%% file: MS2222.tex
\documentclass[]{aa}
\usepackage{astron}
\usepackage{graphics}
\usepackage{rotating}
\usepackage{natbib}
\bibpunct{(}{)}{;}{a}{}{,} 

\input header.tex

\input commande.tex

\begin{document}
\input{aas_macros.tex}

\title{The stellar mass to light ratio in the isolated spiral NGC 4414}
\author{O.~Vallejo\inst{1}, ~J. Braine\inst{1}, A.~Baudry\inst{1}}
\offprints{O.~Vallejo, e-mail: vallejo@observ.u-bordeaux.fr}
\institute{
Observatoire de Bordeaux, UMR 5804, CNRS/INSU, B.P. 89, 
	      F-33270 Floirac, France.} 
\date{Received 20 December 2001/ Accepted 19 February 2002}
\authorrunning{Vallejo et al.}
\titlerunning{The mass to light ratio in NGC 4414}
\abstract{
We present high resolution CO(1-0) interferometric observations and deep 
HST B-V-I images of the flocculent isolated Sc type spiral NGC 4414.  The 
goal is to determine the stellar mass-to-light (M/L) ratio in a galactic disk. 
NGC 4414 is an ideal object for this kind of study, as it is an unperturbed 
object at high galactic latitude with very extended atomic gas (H{\sc i}). 
Many Cepheid light curves were measured in NGC~4414 so its distance is 
known to be about 19.2 Mpc.  NGC~4414 is quite axisymmetric, with no bar and
poorly defined spiral structure, and the center is seen unobscured (no CO, 
HI, H$\alpha$, or thermal dust emission near the nucleus), as in many 
isolated spiral galaxies.  Not only does this result in minimal non-circular 
velocities but also, and this is a key to our success, the central light
profile traces the total mass.  The stars are seen without a dust screen, the
central gas mass is very low (undetected), and we show that the dark matter is
negligible in the central regions.
We have developed an axisymmetric analytical gravitational
potential model to account for the central light (mass) profile, the dynamics 
of the molecular gas in the highly obscured molecular ring, and the stellar 
light profile outside the highly obscured region.  A single dominant disk 
component reproduces the disk dynamics and outer stellar light profile such
that even if other disk components were present they would not affect our 
results.  The contribution of dark matter is constrained by the extremely 
extended HI rotation curve and is small, possibly negligible, at
distances less than 5 -- 7 kpc from the center.  Furthermore, the M/L ratios
we derive are low, about 1.5 in I band and 0.5 in K$'$ band.  The B and V band
M/L ratios vary greatly due to absorption by dust, reaching 4 in the molecular 
ring and decreasing to about 1.6 -- 1.8 at larger radii.  This unequivocally
shows that models, like most maximum disk models, assuming constant M/L ratios
in an optical waveband, simply are not appropriate.  We illustrate this by
making mock maximum disk models with a constant V band M/L ratio.  The key is
having the central light distribution unobscured such that it can be used to
trace the mass. The K' band M/L ratio is virtually constant over the disk,
suggesting that the intrinsic (unobscured) stellar M/L ratio is roughly
constant.  A primitive attempt to determine the intrinsic M/L ratio yields
values close to unity in the B,V, and I bands and slightly below 0.5 in K'.\\ 
}

\maketitle  

\section{Introduction}

The presence and nature of Dark Matter (hereafter DM) is one of if not 
the major problem in modern Astrophysics.  There is evidence for DM
at the scale of individual galaxies, in clusters, and in the universe
as a whole \citep[see reviews by][ and references therein]{Ashman92,sofue01}, 
although a cosmological constant may replace DM at the largest scale.  
Recent results \citep{Alcock98,Renault98}
make it appear unlikely today that all dark matter could be baryonic,
although a large fraction if not all of the DM currently "observed" 
in galaxies could be baryonic.   
Estimating the content and distribution of DM in spiral galaxies
is actually very difficult.  The first essential step to understand the role 
of DM in galaxies is to successfully identify the mass contributions of the 
visible -- stellar and gaseous -- components. That is the subject of this work.

\begin{figure*}
\resizebox{\hsize}{!}{\rotatebox{270}{\includegraphics
{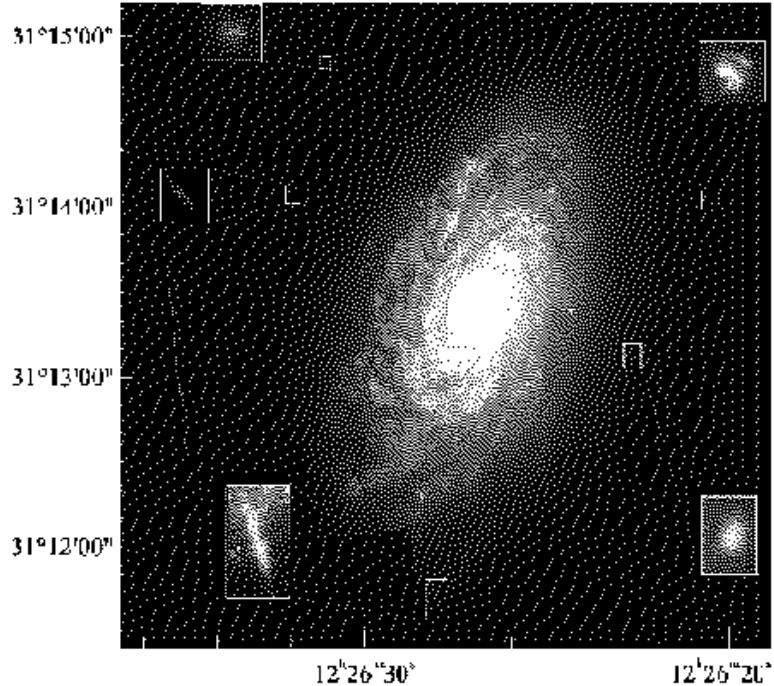}}}
\caption{True color image of the flocculent spiral galaxy NGC 4414 derived 
from HST B, V, and I band images.   Coordinates are J2000.  The dynamic 
range in the image is slightly over $10^4$, 
making the faint emission covering the entire field difficult to see.
Some of the individual background galaxies seen through the disk of NGC~4414 
are highlighted.  They are generally quite red due to the interstellar medium 
in NGC~4414.} 
\end{figure*}

To do this properly, one must measure 
the rotation curve in a spiral with extended atomic 
hydrogen (HI).  The rotation curve must be observed with high spatial 
resolution near the center to avoid beam-smearing effects.  Because 
the visible matter dominates the mass in the central regions of spiral galaxies
(even in Cold Dark Matter, hereafter CDM, simulations), the central
part of the rotation curve can be used to estimate the mass-to-light
(M/L) ratio of the stellar component. 
To minimize stellar 
confusion and galactic absorption, our galaxy of interest should be 
as far as possible from the galactic plane.  The last requirement is 
that the galaxy be at a well known distance.  An Sc type would be 
ideal because they are the most numerous and the mass and luminosity
of the bulge are low compared to the disk, making mass-modeling in 
principle simpler.  The galaxy presented here, NGC~4414, satisfies all
of these criteria. NGC~4414 is an isolated object and its dynamics
are not perturbed by gravitational encounters with other visible galaxies.

\begin{table}[h]
\label{tab:prop}
\begin{center}
\begin{tabular}{llr}
 & & reference \\
\hline
	    $R.A._{2000}$ & $12^h26^{min}27.15^s$ & a \\
	    $Decl_{2000}$ & $+31^\circ 13'24.0"$ & a\\
	    Inclination & $55^\circ$ & a \\
	    Position Angle & $157^\circ$ & a\\
	    Morphological Type & Sc & b \\
	    Optical Diameter $D_{25}$ & 4' & c\\
	    Distance  & 19.2 $\pm$ 2 Mpc & d\\
	    Systemic Velocity & 716 km s$^{-1}$ & e\\
\hline
\end{tabular}
\vskip 1pc
\caption[]{Basic properties of NGC 4414.  \\
a: this paper, position of nucleus determined from HST observations of northern field.
b: \citet{rc3}.
c: this paper.
d: \citet{Turner98,Thim00}.
e: \citet{Braine_n4414a}}
\end{center}
\end{table}

One of the most important questions to resolve is the value of the 
stellar mass-to-light ratio (M/L) of the galactic disk.  In order to 
model rotation curves of spiral galaxies, the stellar light distribution
in a given band is often scaled by the highest M/L ratio possible without
generating higher rotation velocities than are actually observed in the 
inner disk -- the so-called maximum disk model.  
This underlines the need for spatial and spectral resolution.
Carbon Monoxide (CO) is an ideal tracer of the rotation curve in the 
disk because the CO emission comes from the cool dense gas which is naturally
closer to the galactic plane and dynamically cold.  The lines are much 
narrower than the H$\alpha$ (or H$\beta$ etc.) lines and do not suffer from extinction.  The molecular gas, traced by
the CO lines, has higher column densities than the HI and can be observed 
at higher angular resolution.  We present high resolution
observations of the CO lines obtained with the IRAM interferometer. Earlier interferometric CO
observations of NGC 4414 were presented by \citet{Sakamoto96} and \citet{Thornley97b}. 
The large scale molecular emission has been discussed in 
\citet{Braine_n4414a,Braine_n4414b,Braine_n4414c}, in which the $\ratioo$ 
ratio and its variation with galactocentric radius have been estimated.
The H$\alpha$ emission, tracing star formation and ionized gas, was measured
by \citet{Pogge89}.

Our goal is to strongly constrain the M/L ratio in the disk of NGC~4414
using a combined interferometric CO (inner part) and HI (outer part) 
rotation curve.  NGC~4414 was part of the Hubble Space Telescope (HST)
Cepheid key project \citep{Turner98} and thus very deep and detailed images are 
available in the B, V, and I bands. In addition  the distance is known to within 
about 10\%.  We have chosen the I-band ($\sim 0.8 \mu$m) as our reference 
stellar brightness because the longer wavelength should better trace the 
stellar mass and is less affected by extinction than the B or V bands.

An absolute distance D is a necessary parameter to estimate the M/L ratio.
The luminosity is determined as $L = S \times 4\pi \times D^2$, where S represents a flux, whereas the mass M is
proportional to the galactocentric distance: $M \propto R \times V^2
\propto \theta \, D \times V^2$ where $\theta$ is the angular distance 
to the point whose rotation velocity is $V$.  
Thus, $\frac{M}{L} \propto D^{-1}$ so D must be known.

In Section 2 we present the new observational data - the HST B-V-I data and 
the CO(1-0) Plateau de Bure observations. In Section 3 we describe our 
gravitational potential model, based on the interferometric CO observations 
and the HST light profiles, to calculate (Sect. 4) the M/L ratio in the B, 
V, and I bands after subtraction of the gaseous mass.  Our results are compared 
to the rather rare literature estimates of stellar M/L values and to the
so-called ``maximum disk'' method.  

In a companion paper we discuss the 
dark matter distribution in NGC 4414 based on our knowledge of the underlying 
visible mass and on various spherical and disk DM distributions, particularly
those proposed by \citet{Navarro96} and \citet{Moore98b,Moore99} derived from 
simulations of galaxy formation in a CDM (or $\Lambda$CDM) universe.  
Whether Modified Newtonian Dynamics \citep[MOND, ][]{Milgrom83a}
can explain the rotation curve of NGC~4414 without DM is also addressed in the
companion paper.

\section{Observations and Data Reduction}
\subsection{HST BVI Observations}
\subsubsection{Data Reduction}
The optical observations presented in this paper have been retrieved from 
the HST public archive (see Table 2 and \citet{Turner98}).

\begin{table}[h]
\label{tab:obslog}
\begin{center}
\begin{tabular}{cccc}
\hline
 &Optical & Obs.  & Total Observing    \\ 
 &Band &  Date & Time  \\
\hline
 North & F439W (4311.85 \AA)  & Ap. 1996 & 5000 s \\
 North & F555W (5442.23 \AA)	& Ap. 1996 & 32430 s  \\
 North & F814W (8001.59 \AA)	& Ap. 1996 & 10230 s  \\
 South & F439W (4311.85 \AA)  & Ap. 1999 & 2080 s  \\
 South & F555W (5442.23 \AA)	& Ap. 1999 & 1600 s  \\
 South & F814W (8001.59 \AA)	& Ap. 1999 & 1600 s  \\
\hline
\end{tabular}
\vskip 1pc
\caption[]{Dates and total observing times per filter for the northern and southern parts of 
NGC~4414.}
\end{center}
\end{table}

We used WFPC2 automatic calibrated data corrected for ADC errors, bias and 
superbias levels, superdarks, obturation and pixel surface. The classical 
reduction procedure was applied using the IRAF software.
We first corrected all the images for hot pixels, then cosmic rays were 
eliminated by combining consecutive exposures of a same field.

After this corrective stage, a mosaic of the 4 fields could be retrieved, 
or we could treat each field separately.  Our goal was to obtain a complete 
mosaic of the galaxy, with the North (1996 HST Key Project) and South (1999
Hubble Heritage Project) part processed independently.  

The next step was to estimate the shift between each image before adding 
them. The procedure is quite simple, only requiring selection of a few stars 
in each band.  Because we calculate the shift and the rotation simultaneously, 
the stars must be sufficiently bright and homogenously spread over the whole image. 
This calculation was made using the GEOMAP procedure which models a
polynomial transfer function between the reference image and each of the 
other images.  All images in each band were then combined and rotated
to obtain the B-V-I mosaics. (see Fig. 1)

\subsubsection{Luminosity in the outer part of the galaxy}

A bias problem appeared in the HST photometry data as the 
surface brightness observed far away from the center of the galaxy was 
abnormally strong. Such a luminosity could not have a physical
meaning as it was more than an order of magnitude higher than expected. 
Because of the size of the galaxy it was not possible to calculate the 
sky background directly from the HST images.

Optical observations of numerous spiral galaxies, including NGC 4414, were 
made by \citet{Heraudeau96} in order to determine reliable luminosity profiles.  
Conveniently, they deduced luminosity profiles in the B, V and I bands.  
We chose to calculate the surface brightness $\mu$ (in $mag.sec^{-2}$) 100 
arcseconds from the center on the major axis of the galaxy.  We subtracted a 
constant value to set the brightness in each band of the HST data at $100''$ 
to the surface brightness given by \citet{Heraudeau96}.  The uncertainties 
for their photometry are quite low and the 
HST brightness profile after subtraction shows no odd bumps nor negative values.  
At higher brightness this problem becomes irrelevant and the agreement 
between the \citet{Heraudeau96} and the uncorrected HST data is good.  

\begin{figure}[h]
\begin{center}
\includegraphics[angle=270,width=9cm]{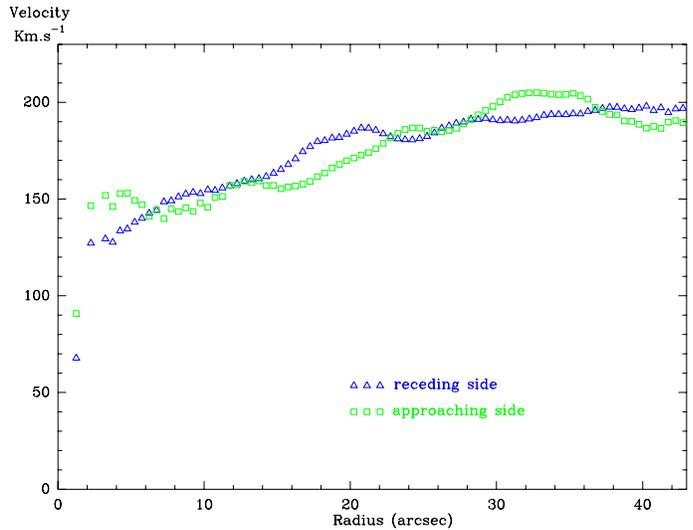}
\caption{CO(1-0) rotation curve near the major axis as derived from the 
interferometer data and excluding points further than 5$^\circ$ 
from the major axis.  The observed velocities shown here are not 
corrected for inclination.} 
\label{fig:courbrot}
\end{center}
\end{figure}

The conversion of instrumental magnitude to physical magnitude in the V 
band is immediate because it corresponds to the Johnson-Morgan 
photometric system \citep{Holtzman95a}.  In the other bands we have to take 
into account the galaxy colors, presented as the colors of specific 
stellar populations.  To identify the most representative
spectral type of stars in the galaxy we compared the B-V and U-B colors of 
NGC~4414, $B-V \approx 0.8$ and $U-B \approx 0.1$, to the colors of the
stellar populations given in Table 3-3 of \citet{Mihalas81}.
Around these values, the corrections remain small for stars of spectral types
between roughly F2 and M0; it is only for the spectral types O5 and M6 that the
corrections can reach 0.1 or 0.2 magnitudes. For G0 and K0 the corrections 
remain nearly constant and we adopt these values (Table 3).

\begin{table}[h] 
\label{tab:corr}  
\begin{center}
\begin{tabular}{l*{4}{c}}
\hline
 & True brightness & Biased brightness & Zero point   \\
 & $mag.sec^{-2}$ & $mag.sec^{-2}$ & Johnson \\\hline
  B-F439W  & $24.15 $& $22.88 $& $+0.55$  \\ 
  V-F555W  & $23.59 $ & $22.27 $ &$+0.00$ \\
  I-F814W  & $22.58 $ & $21.28 $& $-1.30$\\ \hline
\end{tabular}
\vskip 1pc
\caption[]{Surface brightness corrections. Zero points are given in the Johnson-Morgan photometric system, the
value corresponds to average of the corrections from all the spectral types, even if the difference between F2 and M0
are small.}
\end{center}
\end{table}

\subsubsection{Solar luminosity}
To determine the solar luminosity in each WFPC2 optical band we integrated 
a synthetic solar spectrum \citep{Hauschildt99} over the bandpass of each 
HST filter.  The results are given in Table 4.  
We can now express the surface brightness for each band from the flux per 
unit wavelength or per unit frequency.

\begin{table}[h]
\begin{center}
\label{tab:solar}
\caption[]{Solar luminosity in each HST band}
\begin{tabular}{cccc}\hline
  Filter & Pivot   & Bandwidth        & Solar  \\ 
         &     Wavelength &                  &     Luminosity     \\ \hline
	F &  $\lambda_c$      & $\Delta\lambda$  &  $L_\odot$ ($erg.s^{-1}$)     \\ 
	\hline
  439W  & 4311.85 \AA& 404.60 \AA&1.97592 $\times 10^{32}$ \\ 
  555W  & 5442.23 \AA& 1044.65 \AA&5.66221 $\times 10^{32}$ \\ 
  814W  & 8001.59 \AA& 1297.07 \AA&4.27211 $\times 10^{32}$ \\ 
  \hline
\end{tabular}
\end{center}
\end{table}

\subsection{IRAM Plateau de Bure interferometer CO(1-0) observations}
We observed the $J=1\rightarrow0$ emission line of the ${}^{12}CO$ in 
NGC~4414 using the IRAM Plateau de Bure interferometer
between May 1996 and April 1997. The five antennas of the array were equiped with 
dual-band SIS receivers yielding SSB receiver temperatures around 40 K at 
the observing frequency.  The spectral correlator backend was centered 
at 114.996 GHz with a total bandwith of 360 MHz. 
The system temperature varied from about 300K to 400K (Table 5).

\begin{table*}[t]
\small
\centering
\caption{CO(1-0) Observing Parameters with the IRAM Interferometer}
\begin{tabular}{lrrrrr}
\hline
\hline
        &&&&&\\
	Obs. Date & 1996 May 5& 1996 Oct. 27& 1996 Nov. 15&1996 Dec. 15 &1997 Apr. 6\\
	Hour Angle  &-0.9 to 2.9 &-1.0 to 4.9 &-0.6 to 3.8 &-3.8 to 0.6 &-5.0 to -0.8 \\
	Configuration&D&D&C&C&D\\
	Calibrator Flux density (Jy)  &3C273, 23.7 &1156+295, 1.6 &3C273, 24.4
	 &3C273, 22.0 &3C273, 23.0  \\
	&&&&&\\
	\hline
	&&&&&\\
	Primary Beam& \multicolumn{5}{c}{43.8 "} \\
	Restoring Beam&\multicolumn{5}{c}{3.28 " $\times$ 3.00 "} \\
	Velocity Resolution &\multicolumn{5}{c}{6.5 km s$^{-1}$}  \\
	Total Bandwidth &\multicolumn{5}{c}{360 MHz}  \\
	Typical T$_{SYS}$ & \multicolumn{5}{c}{300-400 K}\\
	&&&&&\\
\hline
\hline
\end{tabular}
\label{tbl:pdbi}
\end{table*}

The phase and amplitude calibrations were done 
in the antenna based mode using the strong quasars 3C273 and  1156+295 (Table 5). 
The data were corrected for residual atmospheric phase fluctuations. 
The flux density of the calibrators was determined from IRAM measurements and used
 to derive the absolute flux density scale in our map. 
The accuracy of the flux density scale was around 10\%.
\begin{figure}[h]
\begin{center}
\includegraphics[angle=270,width=9cm]{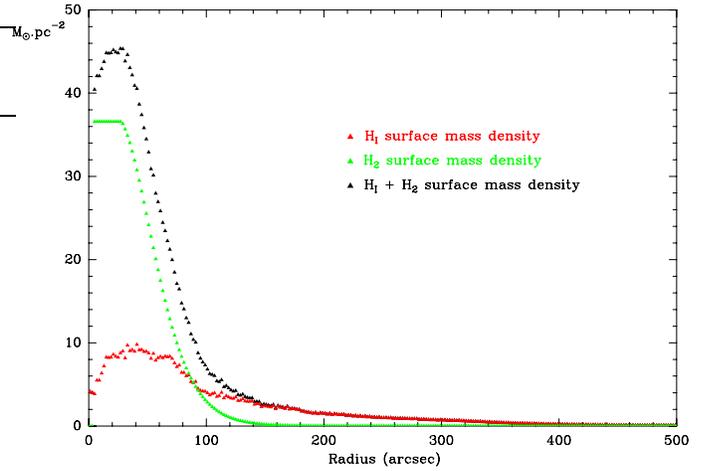}
\caption{HI, H$_2$, and HI + H$_2$ gaseous surface mass densities corrected
to face-on.  He is not included in the figure to make it more easily comparable
to previously published data. } 
\end{center}
\end{figure}

A mosaic map was made from our observations of six fields 
separated by 21$''$ and covering the CO-bright disk of the galaxy. The image
 reconstruction was done using the GILDAS software. 
 Several visibility tables were created each 2.5
MHz wide; the channel separation was thus 6.5 km.s$^{-1}$.
Natural weighting was used with no taper to Fourier transform the visibilities
into images.

The six maps were combined to a single dirty mosaic (256 $\times$ 256 pixels 
with 0.45" sampling), then deconvolved and corrected from primary beam 
attenuation.  The final dirty mosaic was cleaned with the CLARK method 
and restored with a 3.28" $\times$ 3.0" gaussian beam (PA = $35^\circ$). 
To reduce contamination of adjacent fields in the cleaning phase due to 
border effects, each field was truncated where the primary beam power reaches 
a level of 10 \%.

\begin{table}[h]
\label{tab:positions}
\begin{center}
\begin{tabular}{crrr}
\hline
  Cardinal Position & R.A. Offset & Dec. Offset \\
\hline
  North & -11.8 "&25.4 "  \\
  North-East & -13.6 "&5.0 "   \\
  North-West    &5.0 " &13.6 "  \\
  South-East & -5.0 "& -13.6 " \\
  South-West    & 13.6 "& -5.0 " \\
  South & 11.8 "& -25.4 "\\  
\hline
\end{tabular}
\vskip 1pc
\caption[]{Pointing centers of each of the six fields observed as part of the
CO(1--0) mosaic with respect to $12^h26^{min}27.09^s$, $+31^\circ 13'22.0"$
(J2000).}
\end{center}
\end{table}

\subsubsection{Rotation curve}

The rotation velocity of the molecular gas was calculated from 
the first moment of the Plateau de Bure interferometer CO(1--0) data cube (Fig. 2).
Rather than using all channels in our calculations of the velocity at each 
position, a process which brings in noise because the signal is 
typically only present in a few channels at any given position, we used
the fact that the rotation velocities were roughly known in advance.

The circular velocity $V_m$ given by our initial mass model gives us the 
approximate central velocity at each position and we have integrated the data over a total
of 21 channels (136 km s$^{-1}$) surrounding the model velocity. 
\begin{equation}
 V(x,y) = \frac{\sum\limits_{i=V_{m}-10\atop S_i \geq S_0}^{i=V_{m}+10} V_i
 S_i}{\sum\limits_{i=V_{m}-10\atop S_i \geq S_0}^{i=V_{m}+10} S_i} 
\end{equation}
The validity of this method was checked by looking at a very large number of
spectra "by hand".  The linewidths are much smaller than 21 channels so 
we are sure that we have included all possible velocities where emission 
could be present and thus that our moment calculation is not biased towards
the model.  In the central regions where the line widths are broad we have 
integrated over a much larger range. 

\begin{figure}[h]
\begin{center}
\includegraphics[angle=270,width=9cm]{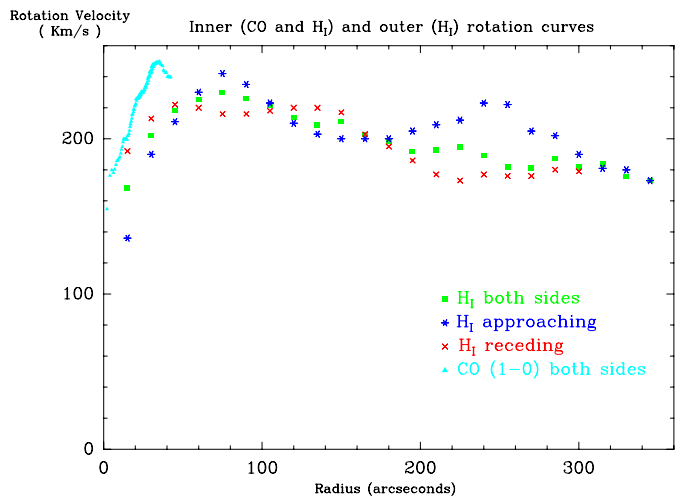}
\caption{CO(1-0) interferometric rotation curve as above but corrected for
inclination shown with low-resolution HI rotation curve from 
\citet{Thornley97b}.} 
\label{fig:h1cocourb}
\end{center}
\end{figure}

\subsection{The large-scale gaseous component}

Before comparing the model mass density with the stellar luminosity 
profiles, we should subtract all of the visible non-stellar mass; namely the gas.
In Paper II, we presented CO(1--0) and CO(2--1) observations over 
almost the entire optical disk of NGC~4414.  $^{13}$CO observations 
were also made over a large part of the disk in both lines.
Using the CO(${{2-1} \over {1-0}}$) line ratio to estimate the gas 
excitation temperature, we used the Virial theorem to estimate the
$\ratioo$ factor to convert CO intensities into H$_2$ mass densities.
As explained in detail in \citet{Braine_n4414b}, our estimates were 
also supported by the $^{13}$CO and 1.2mm thermal dust emission maps.  
We use these results, corrected slightly by the results of our ISO 
LWS observations \citep{Braine_n4414c}, to derive the molecular gas 
surface density.  The total molecular gas mass (including the He in 
the molecular clouds) is about $6.5 \times 10^9 \Mo$.

To determine the atomic gas surface density, we have re-reduced the 
\citet{Thornley97b} HI observations made with the VLA in the C and D 
configurations.  Using uniform weighting the C configuration data
were Fourier transformed and {\sc cleaned} retaining high angular
resolution, $\theta_{\rm fwhp} = 13.1'' \times 12.5''$.  The D configuration data  
were naturally weighted to obtain high surface brightness sensitivity.
The C and D configuration data were combined in the image plane in a 
flux-conserving way to maintain high angular resolution where the
signal is strong and retain the sensitivity to low surface brightness 
features in the outer regions where angular resolution is not as important. 
A very simple method was used: when the brightness of the {\it low} 
resolution map exceeded a certain value, the output map was a linear 
combination 
\begin{equation}
S = (1-\eta) \times C + \eta \times D
\end{equation}
for each pixel 
where $\eta=1$ below the threshold and $\eta = exp^{-aN_{\rm D}}$
where C and D are the C and D array maps and $N_{\rm D}$ is the HI
column density of the pixel in the D configuration map and $a$ is a 
constant chosen such that the C array dominates ($\eta \sim 0.1$) in the
highest brightness regions.  The maximum HI column density in the C array 
map is $3 \times 10^{21}$cm$^{-2}$ and the total mass of the atomic 
gas (HI plus associated He) is $8.4 \times 10^9 \Mo$.  The total 
neutral gas mass (HI + H$_2$ and associated He) is then $1.5 \times 10^{10} \Mo$, which we have 
taken to be the total gaseous mass, neglecting the likely insignificant 
mass of ionized gas.

The HI and H$_2$ column density maps were then used to construct the 
radial surface density profile shown in Fig 3.  With the gaseous surface 
density known at each radius, we could easily subtract its contribution 
to the total mass.  The high-resolution Plateau de Bure data are not used
here because such high angular resolution is not necessary for azimuthally
averaged surface densities while it is important not to resolve out flux
as is the case for interferometric CO maps of galaxies.

\section{Gravitational Potential Model}

In this section we describe the axisymmetric part of the 
gravitational potential of the flocculent galaxy NGC 4414. Axisymmetry is a 
good approximation given the morphology of NGC 4414. The asymmetries 
are relatively small, reaching 20\% in the K$'$ band image of 
\citet{Thornley97a} over only a very small fraction of the disk; this justifies our approach.  

To characterize the gravitational potential, we have used the
\citet{Miyamoto75} potential which is a particularly flexible and simple
{\it analytic} potential-density pair. We compared with an exponential disk
which allows us to retrieve the rotation curve as well as our model, but 
not better. However the exponential disk model has two dimensions,
whereas our model is in three dimensions and can mimic an exponential disk 
or more spherical forms. 

The Miyamoto-Nagai (hereafter M-N) gravitational potential is:

\begin{equation}
\Phi (R,z) = \frac{-GM}{\sqrt{R^2+(a+\sqrt{z^2+b^2})^2}} 
\end{equation}

\begin{figure}[h]
\begin{center}
\includegraphics[angle=270,width=9cm]{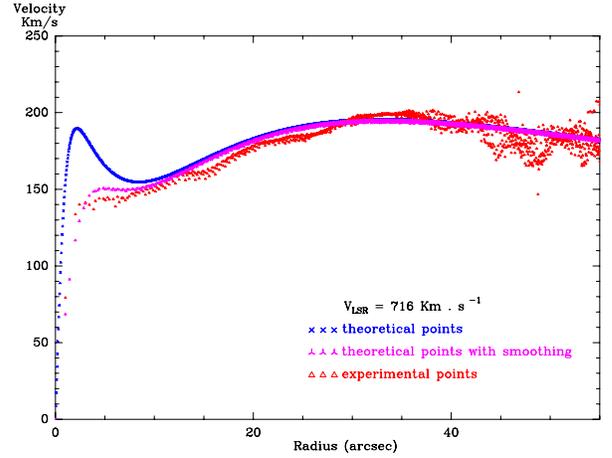}
\caption{Variation of the circular velocity in the galaxy NGC 4414 with 
respect to radius and comparison with interferometric CO(1--0) 
observational data.  The velocities are sky velocities not corrected for
inclination.} 
\label{fig:courbrot}
\end{center}
\end{figure}

The M-N potential is free from any singularities and composed of elementary functions. 
The linearity of Poisson's equation allows us to sum several 
components correponding to different regions of the galaxy, such as the nucleus
and disk.
\begin{eqnarray}
\Phi (R,z) & = & \Phi_1 (R,z) + \Phi_2 (R,z) + ... \nonumber \\
 & = & \Sigma \frac{-GM_i}{\sqrt{R^2+(a_i+\sqrt{z^2+b_i^2})^2}} 
\end{eqnarray}

$a=0$ and $b=0$ correspond to the spherical Plummer potential and the thin-disk Kuzmin potential cases,
respectively. To estimate the mass M and to calculate the density $\rho(R,z)$,
the Laplace equation has to be solved:

\begin{equation}
 \frac{1}{R} \diffpar{}{R} \left(R\diffpar{\Phi}{R}\right) + 
 \diffparcar{\Phi}{z} = 4 \pi G \rho(R,z) 
\end{equation}

Yielding

\begin{equation}
\rho(R,z) = \frac{b^2 M}{4 \pi} \frac{aR^2 + (a + 3 \sqrt{z^2 + b^2})
\left(a + \sqrt{z^2 + b^2}\right)^2}{\left(z^2 + b^2\right)^{\frac{3}{2}} 
\left(R^2 + \left(a + \sqrt{z^2 +b^2}\right)^2\right)^{\frac{5}{2}}} 
\end{equation}

The parameters for this kind of model are $a$, $b$ and $M$, where $a$ is 
the dimension of the galaxy (or given component of the galaxy), $b$ the  
thickness, and $M$ the mass of the galaxy or component described by $a$ 
and $b$. The ratio $\frac{a}{b}$ is a measure of the flatness of the component which becomes a sphere
for $\frac{a}{b} = 0$.

\begin{figure}[h]
\begin{center}
\includegraphics[angle=270,width=9cm]{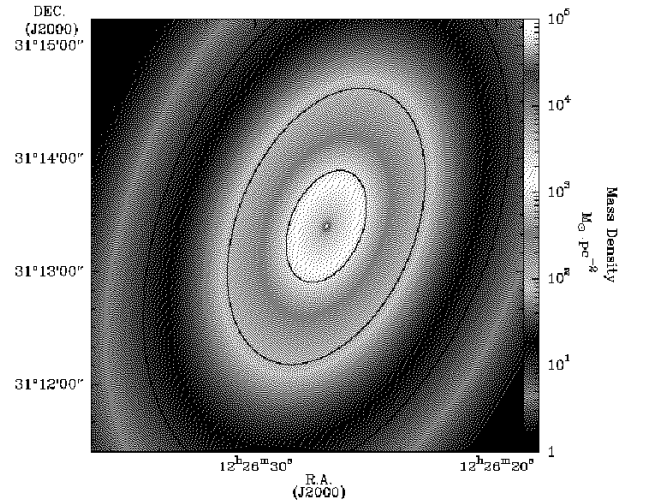}
\caption{Projected surface density of the Miyamoto-Nagai model 
after projection to the same inclination and position angle as NGC 4414.
The ellipses represent circles inclined by 55$^\circ$.} 
\label{fig:densurf}
\end{center}
\end{figure}

The in-plane circular velocity $v_{c} (R)$ of a single component is then given by:
\begin{equation}
v_{c} (R,z=0) = \frac{R \sqrt{GM}}{\left(R^2 + \left(a + b\right)^2\right)^{\frac{3}{4}}}
\end{equation}
or, for several components $\Phi_i$ as in all real cases:
\begin{equation}
v_{c}^2 (R) = \sum\limits_{i} \left(R\diffpar{\Phi_i}{R}\right)_{z=0} = \sum\limits_{i} \left[
\frac{R^2 G M_i}{\left(R^2+(a_i+b_i)^2\right)^{\frac{3}{2}}}\right]
\end{equation}

If the disk is now inclined by an angle $\alpha$ from the line of sight, we could integrate the
volumic density $\rho$ along the line of sight, to obtain the surface density. As the volumic
density is proportional to the mass of the component under consideration, the surface density is too.

Our choice of components was guided by the CO(1--0) high-resolution rotation
curve and by the fact that the mass profile should follow the HST {\it I}-band
light in the center, which is relatively unobscured by gas and dust. In the 
outer regions, beyond the interferometer observations, we have been guided by the fact that the
lower resolution observations show a much lower amount of gas and dust.
We chose the {\it I}-band (F814W) surface brightness because: ($a$) the light 
is less biased to the young stellar population, which represents a small 
fraction of the mass in a spiral galaxy, and ($b$) the absorption of light by 
dust is less than in the {\it B} or {\it V}-bands (F439W and F555W). 

Initially, two components were chosen to represent the dynamics of the 
galaxy, a small bulge component 
and a thin disk component for the rest. The very high central peak 
brightness required us to have two central components in the end, even if we 
could not constrain this region dynamically because no molecular gas is 
present in the nucleus.
We would like to stress that the decomposition into M-N components provide mainly convenient {\it analytic} expressions for the potential 
and density following the rotation curve and light profiles in the 
appropriate regions.  The components adopted for the ``best'' model are given
in Table 7.  The central components occupy the region with little or no
gas and dust and follow the stellar brightness.  Virtually all of the visible
mass is in a single component which represents the disk, essentially stellar
but including the gas as well.  A "negative" component with obviously no physical meaning was added              
in order to make the model fit to the stellar luminosity better in the outer disk;
the effect of this component on the velocities within the visible disk is
negligible (less than 5\%) and is of course included in the calculations.

All of the components are disk-like although the light distribution is slightly
``thicker'' in the inner few arcseconds.  The shape of the smallest component
is influenced only by the light because it is so small ($a=0.7''$) that 
at radii $r \ga$ few arcseconds the rotation velocity is affected only by its
mass and not shape.  The two inner components ($a=0.7''$ and $a=4''$)
enable us to fit the steep rise of the rotation curve.  Two M-N components 
are used to model the stellar + gaseous disk but this should really be viewed 
as a single structure of mass $7 \times 10^{10} M_\odot$ whose mass density 
drops off more quickly with radius than a single M-N component ($a=2500$pc).  
This disk simultaneously reproduces the inner rotation curve and the 
luminosity profile at larger radii, where the dust obscuration is known to
be lower.  It also reproduces the observed disk thickness and inclination.
While other, probably more complicated, mathematical formulations could 
certainly be used, they would have to yield the same result.  

To better compare our model to the CO data, we smoothed the model 
rotation curve using the CO synthesized beam parameters, a $3.28" \times 3.0"$ 
gaussian beam with a position angle of $35^\circ$.  The observed and
model rotation curves are shown in Fig. 5.  An image of the projected
surface density of the mass model we have adopted is shown in Fig. 6.  Comparison with the observed HST
light profiles is shown in Fig. 7.

\begin{table}[h]
\label{tab:components}
\begin{center}
\begin{tabular}{crrr}
\hline
  Component & a (pc) & b (pc) & M ($10^9 M_\odot$) \\
\hline
  Nucleus & 70 &20 & 2.5 \\
  Nucleus & 380 & 30 & 4.0  \\
  Stellar+Gaseous Disk  & 2500 & 150 & 86  \\
  Negative component  & 11000 & 500 & 16 \\
\hline
\end{tabular}
\vskip 1pc
\caption[]{NGC~4414 components according to the gravitational potential model of 
Miyamoto \& Nagai (1975).  The factor $\frac{a}{b}$ indicates the flatness of the component.
The total visible mass is about $7.65 \times 10^{10} M_\odot$.
The thickness ($b$) of the negative component is not well constrained.}
\end{center}
\end{table}

At this stage, our model only contains visible mass components and these are 
shown to dominate at radii less than about 10 kpc ($100''$).  This enables
us to determine the mass-to-light ratio within the optical disk (next section).

\section{The M/L ratio}
  
\begin{figure*}[t]
\begin{center}
\includegraphics[angle=270,width=16cm]{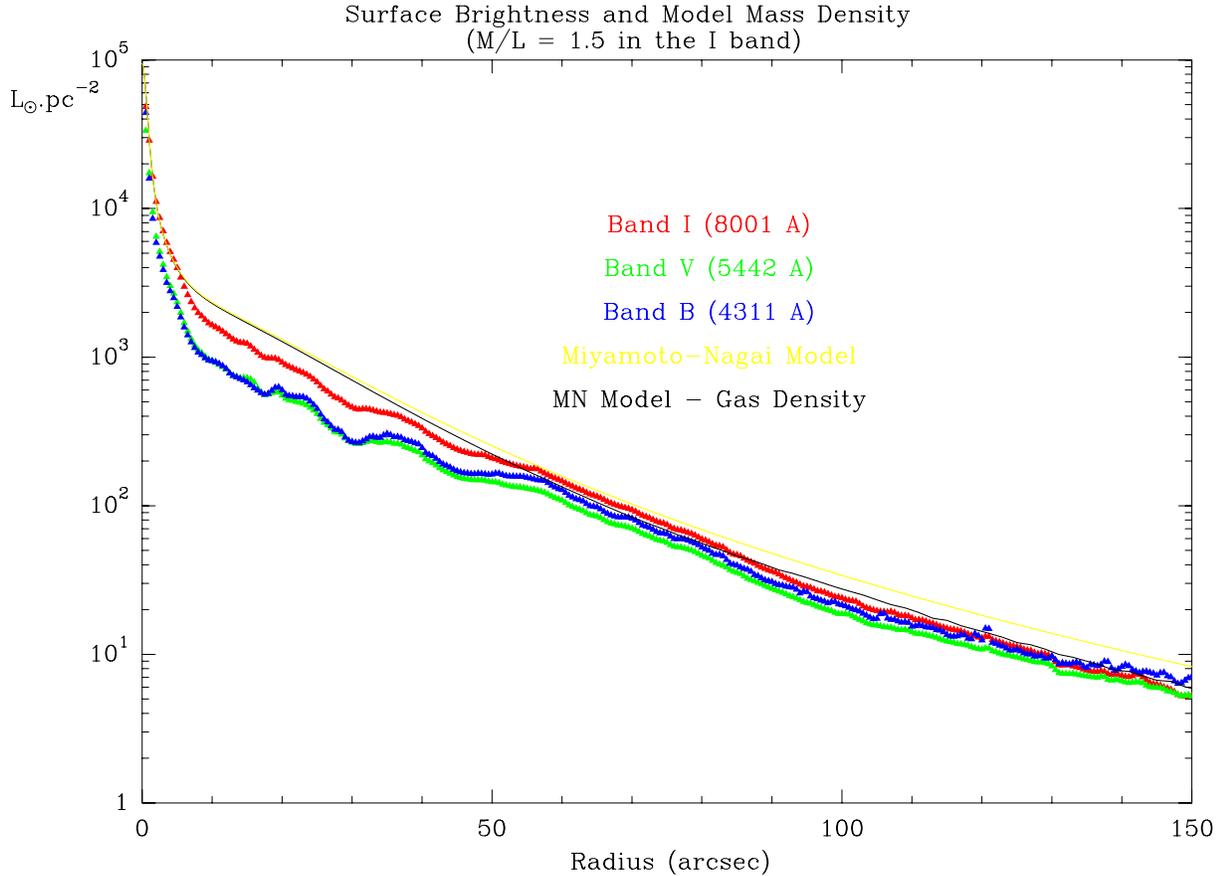}
\caption{B, V, and I band HST luminosity profiles of NGC~4414 with projected
model mass density superposed as a function of radius.  Ordinate is in 
$L_{\odot}$ pc$^{-2}$ where $L_{\odot}$ is the solar luminosity in
each of the B, V, and I bands respectively.  Solid line is the mass model after 
subtraction of the gas mass as described in Sect. 2.3, including Helium, 
and scaled assuming $M/L_{\odot} = 1.5$.  Note how insignificant the gas 
mass is at small radii. }
\label{fig:hstprofil}
\end{center}
\end{figure*}

The mass in the central regions of a spiral galaxy is completely dominated 
by the stellar mass.  For NGC~4414 this can be readily seen by taking the
dark mass necessary to fit the outer point in the HI rotation curve --
roughly $1.5 \times 10^{11} M_\odot$ within the last radius and for a spherical distribution -- and assuming a shape for the halo.
For a rather concentrated ``CDM'' $\rho \propto r^{-1}$ halo \citep{Navarro96},
$\rho(R) = M_{\rm tot}/2 \pi R^3 \, \, \sim 6 \times 10^{-4}M_\odot$pc$^{-3}$
where $R$ is the outer radius of the rotation curve (Fig. 4).  At a radius
of 10pc, the density is then about 2M$_\odot$pc$^{-3}$ and the mass at 
$r < 10$pc is about $10^4$M$_\odot$.  The stellar luminosity is about 
$10^5$ L$_\odot$pc$^{-2}$, yielding a luminosity in the same region of 
$3 \times 10^7$ L$_\odot$.  Even for an M/L of unity, the dark matter
provides less than 0.1\% of the mass in the nucleus.  For $\rho \propto r^{-1.5}$ in the halo
\citep{Moore98b,Moore99}, the dark mass is still only a few percent of the luminous mass.
Clearly, with less centrally concentrated haloes, as is typically suggested by observations
\citep{Cote00,Salucci01,deblok01}, the central contribution of the dark matter will be
even less.

All the conditions are therefore met to derive the stellar M/L ratio in NGC~4414:
$(a)$ the distance is known to within $\sim 10$\%;
$(b)$ the central stellar luminosity is well known and not greatly obscured
(we know this from the CO, HI, and SCUBA observations);
$(c)$ all of the other mass components -- dark matter, neutral gas,
ionized gas, have been shown to be negligible compared to the stellar mass.
For all halo profiles yielding the observed outer rotation curve, the luminous 
mass dominates out to beyond 10 kpc (about $100''$), enabling the stellar 
M/L ratio to be determined as for the center, the gaseous component being
sufficiently well quantified. Following the same calculation as in the previous
paragraph, $\rho \propto r^{-1}$ and $r^{-1.5}$ dark matter profiles yield
dark masses of $3 - 4 \times 10^9 M_\odot$ and $6 - 8 \times 10^9 M_\odot$
within radii of 5 and 7 kpc respectively.  This is much less than the stellar
mass and thus can not significantly affect our estimates of M/L.

The stellar M/L ratio is then the mass model (after subtraction of the small 
gaseous component -- Sect. 2.3), which is dominated by a single component 
of scale length 2.5 kpc, divided by the stellar luminosity and is shown
in Fig. 9.  
In the B and V bands the M/L ratio is between 1.5 and 2 except for 
radii $5'' < r < 50''$ where it reaches about 4 in solar units.  
The same structure is seen in the I band but with lower amplitudes -- 
M/L is about 1.5 reaching slightly over 2.  
It is immediately apparent that the high M/L ratios in the $B$ 
and $V$ bands at radii $5'' < r < 50''$ are due to extinction by the very 
abundant dust (Papers I,II,III and Braine et al. in prep.).  The intrinsic
M/L ratio is about or over unity in each of these bands.  The gas
mass has been subtracted from the mass model before calculating the M/L.
This has no effect in the central regions but becomes important at larger 
radii (compare gas-subtracted and total mass model curves in Fig. 7 for M/L=1.5 in the I band).

The increase in M/L at very small radii is not significant.  The HST images
of the nucleus of NGC 4414 are all, or nearly all, saturated. When combined
and reduced to a common $0.1''$ pixel size, the central flux is underestimated.
To illustrate this, we compare the central portion of the $I$-band profile 
shown in Fig. 7 with a cut across the nucleus in a single relatively short
F606W filter image where the nucleus falls in the PC portion (Fig. 8).  For similar
fluxes at $1''$ from the center, the short exposure has 5 times the central
flux.  Even in this 160 second image, the two central pixels are identical
and only 8\% below the saturated flux of the star about $11''$ away so this
image may be slightly saturated as well.

\begin{figure}[h]
\begin{center}
\includegraphics[angle=0,width=9cm]{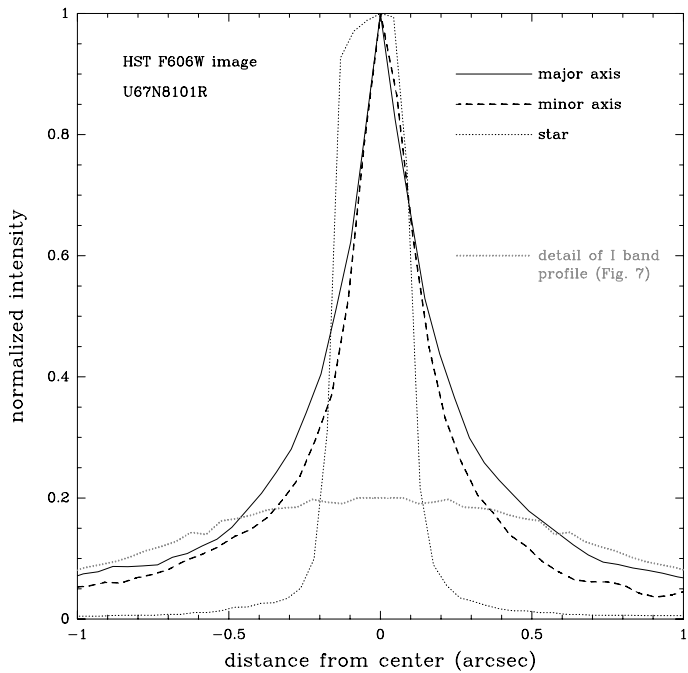}
\caption{Cuts across the major and minor axes of the center of NGC~4414 as
seen in a 160 second exposure with the HST PC in the F606W filter.  For comparison,
a cut across a nearby star and the radial $I$-band profile used in the
mass model are shown.}
\label{fig:hstpcprofil}
\end{center}
\end{figure}

The fact that we see the nucleus of NGC 4414 is an important aid because one is then not tempted to invoke an unknown bulge component
which, although only a small contributor over the whole disk could well
change the shape of the rotation curve in the inner few kpc.  The inner 
region is crucial because it is where we are sure that the stellar mass dominates the dark mass and where our new high-resolution data provide the necessary information.  If the center were covered by dust, it would be
very difficult to estimate the bulge component because beyond the central
region the bulge light would have to be separated from disk light, introducing serious uncertainties.  Seeing the rapid decline in the central
luminosity allows us to determine the central light profile.

\subsection{Comparison with other estimates of M/L}

Few reliable M/L values are available. \citet{Pryor93} studied the mass-to-light ratios for 56 Galactic globular 
clusters using King-Michie dynamical models.  They found a median value of 
the $M/L_V$ ratio of 2.3.  \citet{Dubath97} derived a similar medium value 
for 9 globular clusters in M31. This value is slightly higher than ours 
($M/L_V \approx 2$) in the region where dust is less obstructing, and 
clearly consistent with ours because the stellar population of globular clusters is older
than that of the galactic disk.

In Figure 9 we show K' band mass to light ratio variations in the galactic disk. 
The value is nearly constant, $M/L_{K'} \approx 0.5$, and close to the near 
infrared mass to light ratio given by \citet{Olling01} for the Milky Way disk, 
$0.3 <M/L_{K} < 0.7$ in solar units, and well within the range found by
\citet{Bell01} through population syntheses.  The K' band M/L ratio is low because 
of the contribution of giants which have a particularly strong effect due to 
intrinsically low M/L and their very red colors.  The K' band 
variations are the same as in the optical bands, but with weaker
amplitudes, about 10 \% in the K' band, 40 \% in the I band and reaching factors of 
2 - 2.5 in the V and B bands. These variations result from absorption of the 
starlight by dust so fluctuations appear in the same regions of the galaxy for each band.
The longer wavelength emission suffers less from the extinction so the variations 
decrease with wavelength.

The mass to light ratio we show in Fig 9 can be compared with 
estimates for other spiral disks, typically obtained through mass modeling 
with a constant M/L (usually B band) and yielding results equivalent to 
maximum disk (next subsection).  
\cite{Hoekstra01} find B band M/L ratios from 1.2 to 
5.8, with an average of 3.5 for full-sized spirals.  The extreme values are 
not outliers but the whole range is covered, which intrinsically would not 
be expected for spiral disks.  We suspect that the large range mostly 
reflects the different coverage by dust.

\begin{figure}[h]
\begin{center}
\includegraphics[angle=270,width=9cm]{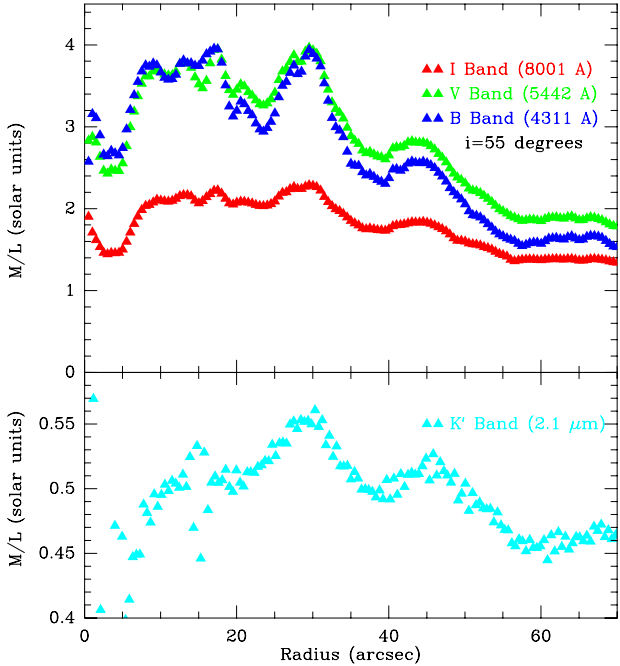}
\caption{The variation of the stellar mass to light ratio 
$(M_{\rm model}-M_{\rm gas})/L_*$ ratio in NGC 4414 as a function of radius.
Note the much greater variation in the B and V bands due to the 
extinction by dust. In the lower panel we show the variation of the 
mass to light ratio in the K' band \citep[observations by ][]{Thornley97b}
with a very small range in M/L in order to show the variations which are 
much smaller than in the shorter wavelength bands.  The central region 
cannot be fairly compared as the $K'$ band resolution is much lower than 
the HST images.} 
\label{fig:msurl}
\end{center}
\end{figure}

\subsection{Comparison with maximum-disk model}

In a maximum disk model \citep{vanalbada86,Palunas00} the mass distribution 
follows the light distribution, 
the M/L ratio being constant along the luminous disk. The disks are called 
maximum because as much mass as possible is attributed to the stellar disk
without generating circular velocities greater than observed.  Typically, 
the band used is either B or V.  To compare 
with our gravitational model we made a maximum disk model by fitting the 
HST V-band luminosity curve. We used four Miyamoto-Nagai components to 
fit the V-band luminosity curve.
A very similar fit is obtained in B band.  We tested two different values 
of  $M/L_{\rm V}$.  $M/L_{\rm V} = 1.7$ fits the inner part of the 
rotation curve at about 5$''$ (a true maximum disk model) and $M/L_{\rm V} = 3.5$ 
fits the peak of the rotation curve at about 35$''$.  To correctly compare
model and observations, we smoothed the 
velocity field with the same parameters as described in Section 3, i.e. 
a $3.28" \times 3.0"$ gaussian beam with a position angle of $35^\circ$. To
complete this study of the maximum disk we included a $\rho \propto 
r^{-1.5}$ spherical halo of Dark Matter. The mass of the halo was calculated 
from the H{\sc i} data taking a $1.6 \times 10^{11} M_\odot $ (M/L = 1.7) or $1.1 \times
10^{11} M_\odot $ (M/L = 3.5) sphere with an outer radius of 33.5 Kpc (i.e. 360 ''). 
A $\rho \propto r^{-1}$ provides a worse fit but the results in this section 
are quite independant of the dark matter halo.

It appears clearly that the velocities generated by a maximum disk (M/L = 3.5) 
are too high from about 5'' to 15'' (500 to 1500 pc) {\it and} that the dark matter 
halo provides no solution to the problem.  The difference is much greater than 
the observational uncertainties.  Lowering the M/L ratio to fit the velocities 
at small radii (M/L = 1.7), say 500 pc, requires a very unphysical dark halo shape.

This shows that a constant M/L ratio, particularly in a short wavelength band
such as B or V, is inappropriate.  Were the nucleus covered or partially covered 
by dust, and thus less luminous, we might not have been able to show that 
M/L cannot be constant.  

\begin{figure}[h]
\begin{center}
\includegraphics[angle=270,width=9cm]{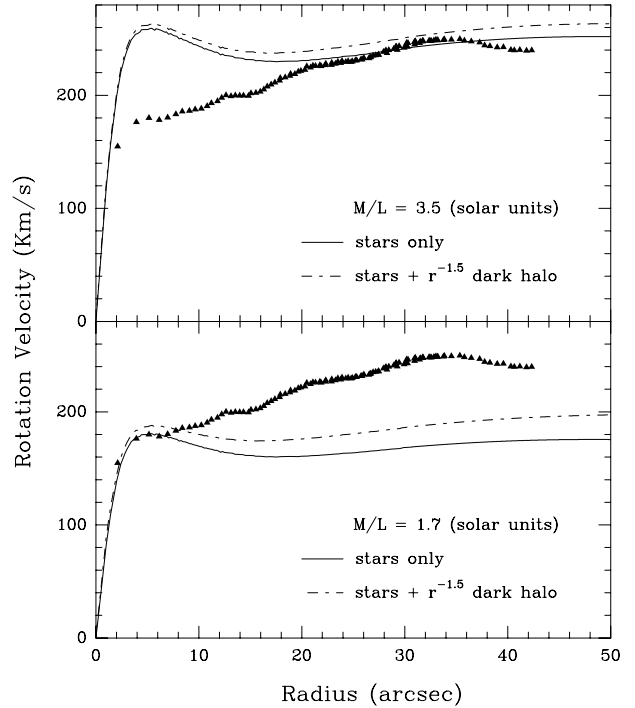}
\caption{Model rotation curve for a maximum disk model compared 
with the high resolution CO(1-0) interferometric data ({\it full triangle}). We present
here a maximum disk model with two possible values of the stellar mass-to-light ratio
to account the rotation velocity curve in the first 10'' (M/L$_{\rm V}$ = 1.7) and farther
than 20'' (M/L$_{\rm V}$ = 3.5). The gas mass is negligible in this zone as can be seen 
in Fig. 7.  We include a $\rho \propto r^{-1.5}$ DM spherical halo which accounts
for the H{\sc i} rotation velocity in the outer part of the galaxy but has
no effect on our conclusions here ({\it dashed lines}).
}
\label{fig:maxdisk}
\end{center}
\end{figure}

\subsection {The intrinsic M/L ratio}

The near constancy of M/L at 2.1 
$\mu$m suggests that the underlying M/L ratio is single-valued over the 
disk.  This is supported by the fact that the increases in K' band M/L 
occur at the same positions as in the B and V bands, with a much lower 
amplitude as expected for extinction, so the overall K' band M/L is very 
close to its value where less extinction is present, about [M/L]$_{\rm K'}
\approx 0.5$.  Below we attempt to estimate the intrinsic M/L ratio of the
stellar population of the disk of NGC 4414.

The M/L ratios in all bands reach roughly constant levels at radii $r \ga 
50''$, outside of the star-forming molecular ring.  Without prior knowledge
of the intrinsic (unreddened) colors, the best way to estimate the 
extinction is through the atomic gas column density.  The HI column density 
at these radii is about $2 \times 10^{21}$cm$^{-2}$ \citep{Thornley97b}, 
such that on average the stellar light will be extincted by about 
$10^{21}$cm$^{-2}$.  From Fig. 8 of \cite{DL84}, we estimate 
$\tau_{\rm ext} \approx 2.3 \times 10^{-26} N_{\rm H}/\lambda$ at the
wavelength of the F814 filter and with $\lambda$ expressed in cm, $N_{\rm H}$ 
in cm$^{-2}$.  This leads to an I band (F814) extinction of about 0.3
in the regions where the I band M/L reaches a plateau at 1.4.

From \cite{DL84} we estimate the extinction ratio between the HST I and V
bands to be about 0.56, between the values given by \cite{rieke85} and
\cite{cohen81}.  Similarly, the HST B to V band extinction ratio should be
about 1.3, again between the values given by \cite{rieke85} and
\cite{cohen81}.  We estimate the ratio between K' and V band extinctions
to be about 0.09, slightly lower than in \cite{rieke85} and in agreement
with \cite{cohen81}.  Using these extinctions to correct the observed 
mass to light ratios leads to M/L ratios of roughly unity in the B,V, and I
bands.  The results are summarized in Table 8.

\begin{table}[h]
\begin{center}
\begin{tabular}{lcccc}
\hline
Band & Observed M/L & $\tau_{\rm ext}$ & $A$ & corrected M/L \\
 & $55'' \ga r \ga 70''$ &&& \\
\hline
B & 1.7 & 0.7 & 0.76 & 0.85 \\
V & 1.9 & 0.54 & 0.58 & 1.1 \\
I & 1.4 & 0.3 & 0.33 & 1.0 \\
$K'$ & 0.46&  0.05 & 0.05 & 0.42 \\
\hline
\end{tabular}
\vskip 1pc
\caption[]{Intrinsic stellar mass to light ratios, corrected for extinction
by dust.  First column specifies the band, where B refers to the 439W filter,
V the 555W, and I the 814W as given in Table 4.  Second col. is M/L before
correction; third col. is extinction expressed as optical depth and in mag.
in col. 4.  Last column gives our estimate of the intrinsic M/L.}
\end{center}
\end{table}

These values should be viewed as upper limits, although they are likely
quite close to the reality, because extinction due to molecular gas
has not been included.  Molecular clouds are smaller denser objects 
that HI clouds and are closer to the galactic plane.  Thus, they may 
very strongly obscure a small fraction of the starlight, causing a minor
overestimate of the mass to light ratio.  At smaller radii
the covering fraction of the clouds is certainly significantly higher.

In the molecular ring, the extinction is roughly 1.6 mag in B, 1.3 in V,
0.9 in I, and 0.2 in $K'$, assuming that the corrected M/L ratios given
above are accurate.  In upcoming papers we plan to determine the extinction 
more accurately by avoiding the azimuthal symmetry hypothesis and studying
the whole galaxy point by point.  Population syntheses may also help to 
constrain the intrinsic colors and M/L ratios, using the values here as
a starting point.  These issues are beyond the scope of this paper.

\section{Conclusion}

This work leads to the following conclusions:
\vskip 1pc
- The mass model is constrained by the light distribution in the central few hundred parsecs
(r $\la$ 4'') and at large radii (r $\ga$ 50'') and by the high-resolution rotation curve at
intermediate radii. A single component dominates the mass over the whole disk so
the agreement between the outer light distribution and rotation curve at smaller
radii is natural.
\vskip 1pc
- Because the central few hundred parsecs are not covered by dust, we are able to
unambiguously show that a constant M/L ratio in the B or V bands is highly
incorrect due to extinction at radii of 5'' $\la$ r $\la$ 50''. A low M/L (1 or 2)
requires an enormous, unphysical, DM contribution at radii greater than 15'' (1.5
kpc).
A higher M/L (e.g. 3) gives far too much mass to the bright central region such
that the calculated rotation velocities are far greater than observed at 5''
$\la$ r $\la$ 20''.
\vskip 1pc
- The M/L ratio, shown in Fig. 9, reaches about 4 in the highly obscured radii 5''
$\la$ r $\la$ 50'' in the B and V bands but is below 2 in the center and the less
obscured outer regions. The I band M/L ratio shows the same structure as in B and
V but only varies between 1.4 and 2. At the longer K' (2.1 $\mu$m) wavelength, the
M/L ratio is almost constant at about 0.5 M$_\odot$/L$_{\odot, K'}$ but again shows the
same absorption regions as in B,V and I.
\vskip 1pc
- The {\it intrinsic} stellar M/L ratio, corrected for extinction, 
is close to unity in the B, V, and I bands, and only 0.4 or so in $K'$ band 
(Table 8).  If the recent \cite{Freedman01} distance is correct, then the M/L 
given here should be raised by 8.5\%, well within the uncertainty of this work.
\vskip 1pc
All of our conclusions deal with the optical disk of NGC~4414 where the visible
matter dominates the rotation velocities, such that the uncertainties linked to
the DM distribution have no effect on these conclusions. In a companion paper we
use our knowledge of the visible mass distribution to constrain the shape of the
dark matter envelope.

\begin{acknowledgements}
We would like to thank Michele Thornley for providing us with her K' band 
image. We would also like to thank the referee, G.D. Bothun, for his helpful remarks.
\end{acknowledgements}

\bibliographystyle{apj}


\end{document}

%% file: header.tex
\def \MH2{M_{\rm H_{2}}}
\def \NH2{N_{\rm H_{2}}}
\def \nh2{n_{\rm H_{2}}}

\def \ratioo{N({\rm H}_2) / I_{\rm CO}}

\def \Mo{{\rm \, M_\odot}}

\def \cm2{{\rm \, cm^{-2}}}

\def\la{\lower.5ex\hbox{$\; \buildrel < \over \sim \;$}}
\def\ga{\lower.5ex\hbox{$\; \buildrel > \over \sim \;$}}

%% file: commande.tex
\def\diffpar#1#2{\frac{\partial #1}{\partial #2}}
\def\diffparcar#1#2{\frac{\partial^2 #1}{\partial {#2}^2}}

%% file: aas_macros.tex
%
%
%


\def\jnl@style{\it}
\def\ref@jnl#1{{\jnl@style#1}}

\def\aj{\ref@jnl{AJ}}                   
\def\araa{\ref@jnl{ARA\&A}}             
\def\apj{\ref@jnl{ApJ}}                 
\def\apjl{\ref@jnl{ApJ}}                
\def\apjs{\ref@jnl{ApJS}}               
\def\ao{\ref@jnl{Appl.~Opt.}}           
\def\apss{\ref@jnl{Ap\&SS}}             
\def\aap{\ref@jnl{A\&A}}                
\def\aapr{\ref@jnl{A\&A~Rev.}}          
\def\aaps{\ref@jnl{A\&AS}}              
\def\azh{\ref@jnl{AZh}}                 
\def\baas{\ref@jnl{BAAS}}               
\def\jrasc{\ref@jnl{JRASC}}             
\def\memras{\ref@jnl{MmRAS}}            
\def\mnras{\ref@jnl{MNRAS}}             
\def\pra{\ref@jnl{Phys.~Rev.~A}}        
\def\prb{\ref@jnl{Phys.~Rev.~B}}        
\def\prc{\ref@jnl{Phys.~Rev.~C}}        
\def\prd{\ref@jnl{Phys.~Rev.~D}}        
\def\pre{\ref@jnl{Phys.~Rev.~E}}        
\def\prl{\ref@jnl{Phys.~Rev.~Lett.}}    
\def\pasp{\ref@jnl{PASP}}               
\def\pasj{\ref@jnl{PASJ}}               
\def\qjras{\ref@jnl{QJRAS}}             
\def\skytel{\ref@jnl{S\&T}}             
\def\solphys{\ref@jnl{Sol.~Phys.}}      
\def\sovast{\ref@jnl{Soviet~Ast.}}      
\def\ssr{\ref@jnl{Space~Sci.~Rev.}}     
\def\zap{\ref@jnl{ZAp}}                 
\def\nat{\ref@jnl{Nature}}              
\def\iaucirc{\ref@jnl{IAU~Circ.}}       
\def\aplett{\ref@jnl{Astrophys.~Lett.}} 
\def\apspr{\ref@jnl{Astrophys.~Space~Phys.~Res.}}
\def\bain{\ref@jnl{Bull.~Astron.~Inst.~Netherlands}} 
\def\fcp{\ref@jnl{Fund.~Cosmic~Phys.}}  
\def\gca{\ref@jnl{Geochim.~Cosmochim.~Acta}}   
\def\grl{\ref@jnl{Geophys.~Res.~Lett.}} 
\def\jcp{\ref@jnl{J.~Chem.~Phys.}}      
\def\jgr{\ref@jnl{J.~Geophys.~Res.}}    
\def\jqsrt{\ref@jnl{J.~Quant.~Spec.~Radiat.~Transf.}}
\def\memsai{\ref@jnl{Mem.~Soc.~Astron.~Italiana}}
\def\nphysa{\ref@jnl{Nucl.~Phys.~A}}   
\def\physrep{\ref@jnl{Phys.~Rep.}}   
\def\physscr{\ref@jnl{Phys.~Scr}}   
\def\planss{\ref@jnl{Planet.~Space~Sci.}}   
\def\procspie{\ref@jnl{Proc.~SPIE}}   

\let\astap=\aap
\let\apjlett=\apjl
\let\apjsupp=\apjs
\let\applopt=\ao

%% file: MS2222.bbl
\begin{thebibliography}{37}
\expandafter\ifx\csname natexlab\endcsname\relax\def\natexlab#1{#1}\fi

\bibitem[{{Alcock} {et~al.}(1998){Alcock}, {Allsman}, {Alves}, {Ansari},
  {Aubourg}, {Axelrod}, {Bareyre}, {Beaulieu}, {Becker}, {Bennett}, {Brehin},
  {Cavalier}, {Char}, {Cook}, {Ferlet}, {Fernandez}, {Freeman}, {Griest},
  {Grison}, {Gros}, {Gry}, {Guibert}, {Lachieze-Rey}, {Laurent}, {Lehner},
  {Lesquoy}, {Magneville}, {Marshall}, {Maurice}, {Milsztajn}, {Minniti},
  {Moniez}, {Moreau}, {Moscoso}, {Palanque-Delabrouille}, {Peterson}, {Pratt},
  {Prevot}, {Queinnec}, {Quinn}, {Renault}, {Rich}, {Spiro}, {Stubbs},
  {Sutherland}, {Tomaney}, {Vandehei}, {Vidal-Madjar}, {Vigroux}, \&
  {Zylberajch}}]{Alcock98}
{Alcock}, C., {Allsman}, R.~A., {Alves}, D., {Ansari}, R., {Aubourg}, E.,
  {Axelrod}, T.~S., {Bareyre}, P., {Beaulieu}, J.-P., {Becker}, A.~C.,
  {Bennett}, D.~P., {Brehin}, S., {Cavalier}, F., {Char}, S., {Cook}, K.~H.,
  {Ferlet}, R., {Fernandez}, J., {Freeman}, K.~C., {Griest}, K., {Grison}, P.,
  {Gros}, M., {Gry}, C., {Guibert}, J., {Lachieze-Rey}, M., {Laurent}, B.,
  {Lehner}, M.~J., {Lesquoy}, E., {Magneville}, C., {Marshall}, S.~L.,
  {Maurice}, E., {Milsztajn}, A., {Minniti}, D., {Moniez}, M., {Moreau}, O.,
  {Moscoso}, L., {Palanque-Delabrouille}, N., {Peterson}, B.~A., {Pratt},
  M.~R., {Prevot}, L., {Queinnec}, F., {Quinn}, P.~J., {Renault}, C., {Rich},
  J., {Spiro}, M., {Stubbs}, C.~W., {Sutherland}, W., {Tomaney}, A.,
  {Vandehei}, T., {Vidal-Madjar}, A., {Vigroux}, L., \& {Zylberajch}, S. 1998,
  \apjl, 499, L9

\bibitem[{{Ashman}(1992)}]{Ashman92}
{Ashman}, K.~M. 1992, PASP, 104, 1109

\bibitem[{{Bell} \& {de Jong}(2001)}]{Bell01}
{Bell}, E.~F. \& {de Jong}, R.~S. 2001, ApJ, 550, 212

\bibitem[{{Braine} {et~al.}(1997){Braine}, {Brouillet}, \&
  {Baudry}}]{Braine_n4414b}
{Braine}, J., {Brouillet}, N., \& {Baudry}, A. 1997, A\&A, 318, 19

\bibitem[{{Braine} {et~al.}(1993){Braine}, {Combes}, \& {van
  Driel}}]{Braine_n4414a}
{Braine}, J., {Combes}, F., \& {van Driel}, W. 1993, A\&A, 280, 451

\bibitem[{{Braine} \& {Hughes}(1999)}]{Braine_n4414c}
{Braine}, J. \& {Hughes}, D.~H. 1999, A\&A, 344, 779

\bibitem[{{Cohen} {et~al.}(1981){Cohen}, {Persson}, {Elias}, \&
  {Frogel}}]{cohen81}
{Cohen}, J.~G., {Persson}, S.~E., {Elias}, J.~H., \& {Frogel}, J.~A. 1981, ApJ,
  249, 481

\bibitem[{C{\^o}t{\'e} {et~al.}(2000)C{\^o}t{\'e}, {Carignan}, \&
  {Freeman}}]{Cote00}
C{\^o}t{\'e}, S., {Carignan}, C., \& {Freeman}, K.~C. 2000, AJ, 120, 3027

\bibitem[{{de Blok} {et~al.}(2001){de Blok}, {McGaugh}, {Bosma}, \&
  {Rubin}}]{deblok01}
{de Blok}, W.~J.~G., {McGaugh}, S.~S., {Bosma}, A., \& {Rubin}, V.~C. 2001,
  ApJL, 552, L23

\bibitem[{{de Vaucouleurs} {et~al.}(1991){de Vaucouleurs}, {de Vaucouleurs},
  {Corwin}, {Buta}, {Paturel}, \& {Fouque}}]{rc3}
{de Vaucouleurs}, G., {de Vaucouleurs}, A., {Corwin}, J.~R., {Buta}, R.~J.,
  {Paturel}, G., \& {Fouque}, P. 1991, in Third reference catalogue of bright
  galaxies (1991)

\bibitem[{{Draine} \& {Lee}(1984)}]{DL84}
{Draine}, B.~T. \& {Lee}, H.~M. 1984, ApJ, 285, 89

\bibitem[{{Dubath} \& {Grillmair}(1997)}]{Dubath97}
{Dubath}, P. \& {Grillmair}, C.~J. 1997, \aap, 321, 379

\bibitem[{{Freedman} {et~al.}(2001){Freedman}, {Madore}, {Gibson}, {Ferrarese},
  {Kelson}, {Sakai}, {Mould}, {Kennicutt}, {Ford}, {Graham}, {Huchra},
  {Hughes}, {Illingworth}, {Macri}, \& {Stetson}}]{Freedman01}
{Freedman}, W.~L., {Madore}, B.~F., {Gibson}, B.~K., {Ferrarese}, L., {Kelson},
  D.~D., {Sakai}, S., {Mould}, J.~R., {Kennicutt}, R.~C., {Ford}, H.~C.,
  {Graham}, J.~A., {Huchra}, J.~P., {Hughes}, S.~M.~G., {Illingworth}, G.~D.,
  {Macri}, L.~M., \& {Stetson}, P.~B. 2001, ApJ, 553, 47

\bibitem[{{Hauschildt} {et~al.}(1999){Hauschildt}, {Allard}, \&
  {Baron}}]{Hauschildt99}
{Hauschildt}, P.~H., {Allard}, F., \& {Baron}, E. 1999, ApJ, 512, 377

\bibitem[{{H\'eraudeau} \& {Simien}(1996)}]{Heraudeau96}
{H\'eraudeau}, P. \& {Simien}, F. 1996, A\&AS, 118, 111

\bibitem[{{Hoekstra} {et~al.}(2001){Hoekstra}, {van Albada}, \&
  {Sancisi}}]{Hoekstra01}
{Hoekstra}, H., {van Albada}, T.~S., \& {Sancisi}, R. 2001, MNRAS, 323, 453

\bibitem[{{Holtzman} {et~al.}(1995){Holtzman}, {Hester}, {Casertano},
  {Trauger}, {Watson}, {Ballester}, {Burrows}, {Clarke}, {Crisp}, {Evans},
  {Gallagher}, {Griffiths}, {Hoessel}, {Matthews}, {Mould}, {Scowen},
  {Stapelfeldt}, \& {Westphal}}]{Holtzman95a}
{Holtzman}, J.~A., {Hester}, J.~J., {Casertano}, S., {Trauger}, J.~T.,
  {Watson}, A.~M., {Ballester}, G.~E., {Burrows}, C.~J., {Clarke}, J.~T.,
  {Crisp}, D., {Evans}, R.~W., {Gallagher}, J.~S., {Griffiths}, R.~E.,
  {Hoessel}, J.~G., {Matthews}, L.~D., {Mould}, J.~R., {Scowen}, P.~A.,
  {Stapelfeldt}, K.~R., \& {Westphal}, J.~A. 1995, PASP, 107, 156

\bibitem[{{Mihalas} \& {Binney}(1981)}]{Mihalas81}
{Mihalas}, D. \& {Binney}, J. 1981, {Galactic astronomy: Structure and
  kinematics /2nd edition/} (San Francisco, CA, W.~H.~Freeman and Co.,
  1981.~608 p.)

\bibitem[{{Milgrom}(1983)}]{Milgrom83a}
{Milgrom}, M. 1983, \apj, 270, 365

\bibitem[{{Miyamoto} \& {Nagai}(1975)}]{Miyamoto75}
{Miyamoto}, M. \& {Nagai}, R. 1975, PASJ, 27, 533

\bibitem[{{Moore} {et~al.}(1998){Moore}, {Governato}, {Quinn}, {Stadel}, \&
  {Lake}}]{Moore98b}
{Moore}, B., {Governato}, F., {Quinn}, T., {Stadel}, J., \& {Lake}, G. 1998,
  ApJL, 499, L5

\bibitem[{{Moore} {et~al.}(1999){Moore}, {Quinn}, {Governato}, {Stadel}, \&
  {Lake}}]{Moore99}
{Moore}, B., {Quinn}, T., {Governato}, F., {Stadel}, J., \& {Lake}, G. 1999,
  MNRAS, 310, 1147

\bibitem[{{Navarro} {et~al.}(1996){Navarro}, {Frenk}, \& {White}}]{Navarro96}
{Navarro}, J.~F., {Frenk}, C.~S., \& {White}, S.~D.~M. 1996, ApJ, 462, 563

\bibitem[{{Olling} \& {Merrifield}(2001)}]{Olling01}
{Olling}, R.~P. \& {Merrifield}, M.~R. 2001, \mnras, 326, 164

\bibitem[{{Palunas} \& {Williams}(2000)}]{Palunas00}
{Palunas}, P. \& {Williams}, T.~B. 2000, \aj, 120, 2884

\bibitem[{{Pogge}(1989)}]{Pogge89}
{Pogge}, R.~W. 1989, ApJS, 71, 433

\bibitem[{{Pryor} \& {Meylan}(1993)}]{Pryor93}
{Pryor}, C. \& {Meylan}, G. 1993, in ASP Conf. Ser. 50: Structure and Dynamics
  of Globular Clusters, 357+

\bibitem[{{Renault} {et~al.}(1998){Renault}, {Aubourg}, {Bareyre}, {Brehin},
  {Gros}, {Lachieze-Rey}, {Laurent}, {Lesquoy}, {Magneville}, {Milsztajn},
  {Moscoso}, {Palanque-Delabrouille}, {Queinnec}, {Rich}, {Spiro}, {Vigroux},
  {Zylberajch}, {Ansari}, {Cavalier}, {Moniez}, {Beaulieu}, {Ferlet}, {Grison},
  {Vidal-Madjar}, {Guibert}, {Moreau}, {Maurice}, {Prevot}, {Gry}, {Char},
  {Fernandez}, \& {The EROS Collaboration}}]{Renault98}
{Renault}, C., {Aubourg}, E., {Bareyre}, P., {Brehin}, S., {Gros}, M.,
  {Lachieze-Rey}, M., {Laurent}, B., {Lesquoy}, E., {Magneville}, C.,
  {Milsztajn}, A., {Moscoso}, L., {Palanque-Delabrouille}, N., {Queinnec}, F.,
  {Rich}, J., {Spiro}, M., {Vigroux}, L., {Zylberajch}, S., {Ansari}, R.,
  {Cavalier}, F., {Moniez}, M., {Beaulieu}, J.-P., {Ferlet}, R., {Grison}, P.,
  {Vidal-Madjar}, A., {Guibert}, J., {Moreau}, O., {Maurice}, E., {Prevot}, L.,
  {Gry}, C., {Char}, S., {Fernandez}, J., \& {The EROS Collaboration}. 1998,
  \aap, 329, 522

\bibitem[{{Rieke} \& {Lebofsky}(1985)}]{rieke85}
{Rieke}, G.~H. \& {Lebofsky}, M.~J. 1985, ApJ, 288, 618

\bibitem[{{Sakamoto}(1996)}]{Sakamoto96}
{Sakamoto}, K. 1996, ApJ, 471, 173+

\bibitem[{{Salucci}(2001)}]{Salucci01}
{Salucci}, P. 2001, MNRAS, 320, L1

\bibitem[{{Sofue} \& {Rubin}(2001)}]{sofue01}
{Sofue}, Y. \& {Rubin}, V. 2001, ARA\&A, 37, 137

\bibitem[{{Thim}(2000)}]{Thim00}
{Thim}, F. 2000, in ASP Conf. Ser. 203: IAU Colloq. 176: The Impact of
  Large-Scale Surveys on Pulsating Star Research, 231--232

\bibitem[{{Thornley} \& {Mundy}(1997{\natexlab{a}})}]{Thornley97a}
{Thornley}, M.~D. \& {Mundy}, L.~G. 1997{\natexlab{a}}, ApJ, 484, 202

\bibitem[{{Thornley} \& {Mundy}(1997{\natexlab{b}})}]{Thornley97b}
---. 1997{\natexlab{b}}, ApJ, 490, 682

\bibitem[{{Turner} {et~al.}(1998){Turner}, {Ferrarese}, {Saha}, {Bresolin},
  {Kennicutt}, {Stetson}, {Mould}, {Freedman}, {Gibson}, {Graham}, {Ford},
  {Han}, {Harding}, {Hoessel}, {Huchra}, {Hughes}, {Illingworth}, {Kelson},
  {Macri}, {Madore}, {Phelps}, {Rawson}, {Sakai}, \& {Silbermann}}]{Turner98}
{Turner}, A., {Ferrarese}, L., {Saha}, A., {Bresolin}, F., {Kennicutt}, R.~C.,
  {Stetson}, P.~B., {Mould}, J.~R., {Freedman}, W.~L., {Gibson}, B.~K.,
  {Graham}, J.~A., {Ford}, H., {Han}, M., {Harding}, P., {Hoessel}, J.~G.,
  {Huchra}, J.~P., {Hughes}, S.~M.~G., {Illingworth}, G.~D., {Kelson}, D.~D.,
  {Macri}, L., {Madore}, B.~F., {Phelps}, R., {Rawson}, D., {Sakai}, S., \&
  {Silbermann}, N.~A. 1998, ApJ, 505, 207

\bibitem[{{van Albada} \& {Sancisi}(1986)}]{vanalbada86}
{van Albada}, T.~S. \& {Sancisi}, R. 1986, Royal Society of London
  Philosophical Transactions Series, 320, 447

\end{thebibliography}
